\documentclass[preprint,aps,amsmath,amssymb,12pt]{revtex4}

\usepackage{epsfig}
\usepackage{slashed}
\usepackage{graphicx}
\usepackage{multirow,color}
\usepackage{amsmath}
\usepackage{float}
\usepackage{diagbox}
\usepackage{CJK}
\usepackage{color}
\usepackage{xcolor}
\usepackage{times}
\usepackage{subfigure}
\usepackage{bm}
\usepackage{braket}
\usepackage{booktabs}
\usepackage{array}
\usepackage[mathscr]{euscript}
\usepackage{caption}
\usepackage{makecell}

\makeatletter

\newcommand{\Rmnum}[1]{\expandafter\@slowromancap\romannumeral #1@}
\makeatother
\graphicspath{{fig/}}
\begin{document}
\title{Sensitivity of the future $\boldsymbol{e^{-}p}$ collider to the coupling of axion-like particles with vector bosons}
\author{Chong-Xing Yue$^{1,2}$}
\thanks{cxyue@lnnu.edu.cn}
\author{Han Wang$^{1,2}$}
\thanks{wangwanghan1106@163.com}
\author{Xue-Jia Cheng$^{1,2}$}
\thanks{cxj225588@163.com}
\author{Yue-Qi Wang$^{1,2}$}
\thanks{wyq13889702166@163.com}

\affiliation{
$^1$Department of Physics, Liaoning Normal University, Dalian 116029, China\\
$^2$Center for Theoretical and Experimental High Energy Physics, Liaoning Normal University, China
}

\begin{abstract}
Axion-like particles (ALPs) arise naturally in many extensions of the Standard Model (SM). We explore the discovery potential for ALPs of the Large Hadron electron Collider (LHeC) via the $W^{+}W^{-}$ fusion process. For concreteness, both cases of the ALP decaying to muon pairs and $b\overline{b}$ pairs are investigated. Our numerical results show that the LHeC with the center of mass energy of $1.3$ TeV and the integrated luminosity of $1$ ab$^{-1}$ might be more sensitive than the LHC in probing ALPs over a range of masses from a few tens of GeV to $900$ GeV, where the promising sensitivities to the coupling of ALP with $W^{\pm}$ bosons reach nearly $0.15$ TeV$^{-1}$ and $0.32$ TeV$^{-1}$ for the signal processes $e^{-}p\rightarrow\nu_{e}ja(a\rightarrow\mu^{+}\mu^{-})$ and $e^{-}p\rightarrow\nu_{e}ja(a\rightarrow b\overline{b})$, respectively.
\end{abstract}

\maketitle

\section{Introduction}

The Standard Model (SM) of particle physics has been proved to be remarkably successful, explaining almost all relevant data with a handful of parameters. However, there are many reasons to believe that the SM is not an ultimate theory of nature. For example, it can not solve the gauge hierarchy problem~\cite{Feng:2013pwa} and the strong CP problem~\cite{Kim:2008hd}. The discrepancy between the SM prediction for the muon anomalous magnetic moment and the experimental result~\cite{Lindner:2016bgg}, as well as the problem of neutrino masses~\cite{Gonzalez-Garcia:2007dlo}, can not be explained either. As a result, there are currently a number of areas where new physics beyond the SM (BSM) may exist. High energy collider experiments provide the primary tools to effectively search for BSM physics at the energy frontier. The lack of new physics at the LHC up to now compels new theoretical ideas to be developed and the exploration for complementarities of the $pp$ machines with other potential future facilities. The Large Hadron electron Collider (LHeC)~\cite{LHeCStudyGroup:2012zhm,AbelleiraFernandez:2012ni,Bruening:2013bga,LHeC:2020van} has been proposed to complement the measurements at the LHC, which combines the superior performance of the $pp$ and $e^{+}e^{-}$ colliders. It yields the projected integrated luminosity of $\mathcal{L}=$ $1$ ab$^{-1}$ at the center of mass energy $\sqrt{s}=1.3$ TeV following the baseline energy of the $60$ GeV electron beam in cooperation with the $7$ TeV proton beam. Due to the clean experimental environment and the prospects for the substantial extension of the kinematic range in deep inelastic scattering, the LHeC has the potential to find new physics. In fact, the axion-like particle (ALP, known as ``$a$") is clearly an interesting BSM scenario that is worthy of being studied at the LHeC.

ALPs are pseudo-Nambu-Goldstone bosons, typically those induced by the spontaneous breaking of one or more global symmetries. These particles are independent of Peccei-Quinn mechanism and enjoy less model constraints than the QCD axion \cite{Peccei:1977hh,Weinberg:1977ma,Wilczek:1977pj,GrillidiCortona:2015jxo}. They are CP-odd scalars and gauge-singlets under the SM, which are well motivated new physics candidates. The physics case of the ALP has been greatly developed in recent years, and the experimental landscape is rapidly evolving for exploring the accessible regions in the ALP parameter space.
Whereas astrophysics and cosmology impose tight constraints on very light ALPs~\cite{Raffelt:1990yz,Marsh:2015xka}, the most powerful detections of such particles in the MeV-GeV range arise from experiments performing at the precision frontier~\cite{Konaka:1986cb,Belle-II:2020jti,Altmannshofer:2022izm}. A variety of dedicated experiments have been proposed to probe heavier ALPs in terms of their masses and couplings. In addition to phenomenological studies of ALPs focusing on their interactions with gluons or fermions~\cite{Brivio:2017ije,Ghebretinsaea:2022djg,Yue:2022ash}, searches for the ALP coupled to the SM electroweak gauge bosons ($\gamma$, $W^{\pm}$, $Z$) are also available~\cite{Baldenegro:2018hng,Florez:2021zoo,dEnterria:2021ljz,Inan:2020aal,Inan:2020kif,Zhang:2021sio,Yue:2021iiu,Lu:2022zbe,Bao:2022onq,Han:2022mzp}.  The LHC and the future $e^{+}e^{-}$ colliders have been used to detect the ALP via the photon-photon fusion process, as discussed extensively in Refs.~\cite{Baldenegro:2018hng,Florez:2021zoo,dEnterria:2021ljz,Inan:2020aal,Inan:2020kif,Zhang:2021sio}. The production of ALP could also be efficiently produced through the massive vector boson ($W^{\pm}$, $Z$) fusion processes in which the ALP decays into diphoton at the prospective $e^{+}e^{-}$ and $ep$ colliders~\cite{Yue:2021iiu,Lu:2022zbe}, as well as at the muon colliders~\cite{Bao:2022onq,Han:2022mzp}. It is of great interest to detect the ALP at the future $ep$ colliders. Inspired by previous studies, we implement in this work the investigation of the ALP at the LHeC with $\sqrt{s}=1.3$ TeV and $\mathcal{L}=$ $1$ ab$^{-1}$ through the $W^{+}W^{-}$ fusion process, where the ALP decays to fermions.

The paper is structured as follows. After describing the theory framework in Sec. II, we provide a detailed analysis in Sec. III for the possibility of probing ALPs via the $W^{+}W^{-}$ fusion processes $e^{-}p\rightarrow\nu_{e}ja(a\rightarrow\mu^{+}\mu^{-})$ and $e^{-}p\rightarrow\nu_{e}ja(a\rightarrow b\overline{b})$ based on the LHeC detector simulation. Our main results about the projected LHeC sensitivity region on the coupling of ALP with $W^{\pm}$ bosons are summarized in Sec. IV, where we compare our results with those from the LHC.

\section{The theory framework}

Generally, the couplings of ALP with the SM particles can be encoded in the following effective Lagrangian that includes operators with dimension up to five~\cite{Georgi:1986df,Brivio:2017ije,Bauer:2017ris}

\begin{eqnarray}
\begin{split}
\mathcal{L}_{\text{eff}}
	=
    &\frac{1}{2} (\partial^\mu a)(\partial_\mu a) - \frac{1}{2} m_a^2 a^2 + \frac{\partial^{\mu} a}{f_a} \sum_{\substack{\psi=Q_L,\,Q_R, \\\,L_L,\,L_R}} \bar\psi \gamma_\mu X_\psi \psi \\
    &- C_{\tilde{G}} \frac{a}{f_a} G^{i}_{\mu\nu} \tilde{G}^{i\mu\nu} - C_{\tilde{W}} \frac{a}{f_a} W^{i}_{\mu\nu} \tilde{W}^{i\mu\nu} - C_{\tilde{B}} \frac{a}{f_a} B_{\mu\nu} \tilde{B}^{\mu\nu}\text{,}
\end{split}
\end{eqnarray}
where $G^{i}_{\mu\nu}$, $W^{i}_{\mu\nu}$ and $B_{\mu\nu}$ are respectively the $SU(3)_{C}$, $SU(2)_{L}$ and $U(1)_{Y}$ gauge field strengths, while $\tilde{G}^{i\mu\nu}$, $\tilde{W}^{i\mu\nu}$ and $\tilde{B}^{\mu\nu}$ are the corresponding dual field strengths which are defined as $\tilde{V}^{\mu\nu}=\frac{1}{2}\epsilon^{\mu\nu\lambda\kappa}V_{\lambda\kappa}$ $(V \in G$, $W$, $B)$. $X_\psi$ are $3\times3$ Hermitian matrices in flavour space. The ALP mass $m_a$ and the decay constant $f_a$ can be regarded as independent parameters.

Instead of a number of previous articles focusing on the decay channel $a\rightarrow\gamma\gamma$~\cite{Yue:2019gbh,Liu:2021lan,Yue:2021iiu,Wang:2022ock,Lu:2022zbe}, we explore the possibility of probing ALP at the LHeC through the $W^{+}W^{-}$ fusion process with ALP further decaying to fermions in this work. Then the case that ALP couples to  $W^{\pm}$ bosons and fermions should be considered. After electroweak symmetry breaking, the effective Lagrangian Eq.(1) can  give the couplings of ALP to $W^{\pm}$ bosons and fermions
\begin{eqnarray}
\mathcal{L}_{\text {eff }} &\supset& i g_{a\psi} a \sum_{\psi=Q,\,L}  m^{\text{diag}}_{\psi}\, \bar{\psi} \gamma_5 \psi -\dfrac{1}{4}g_{aWW}aW_{\mu\nu}\tilde{W}^{\mu\nu}\text{,}
\end{eqnarray}
where the sum in the first term extends over all fermion mass eigenstates. $g_{a\psi}$ is the coupling coefficient for the effective ALP-fermion interaction and $m^{\text{diag}}_{\psi}$ is the fermion mass matrix. The interaction of ALP with $W^{\pm}$ bosons comes from the contribution of the fifth term in Eq.(1). The coupling strength of such interaction is depicted by the coupling coefficient $g_{aWW} = \frac{4}{f_a} C_{\tilde{W}}$. The model file with the Lagrangian is produced by \verb"FeynRules"~\cite{Alloul:2013bka}.

The flavor bounds of quark and lepton flavor-changing processes focusing on the MeV-GeV mass range of ALP in an effective field theory have been recently explored in Ref.~\cite{Bauer:2021mvw}. The most stringent upper limits on $g_{a\psi}$ at $90$$\%$ C.L. come from Beam Dump experiments for ALP with mass interval between $1$ MeV and $3$ GeV, which are ($3.4\times10^{-5} - 2.9\times10^{-3}$) TeV$^{-1}$~\cite{CHARM:1985anb,Dolan:2014ska}. The constraints on $g_{a\psi}$ are much looser in higher mass region of ALP.

\begin{figure}[H]
\begin{center}
\subfigure[]{\includegraphics [scale=0.5] {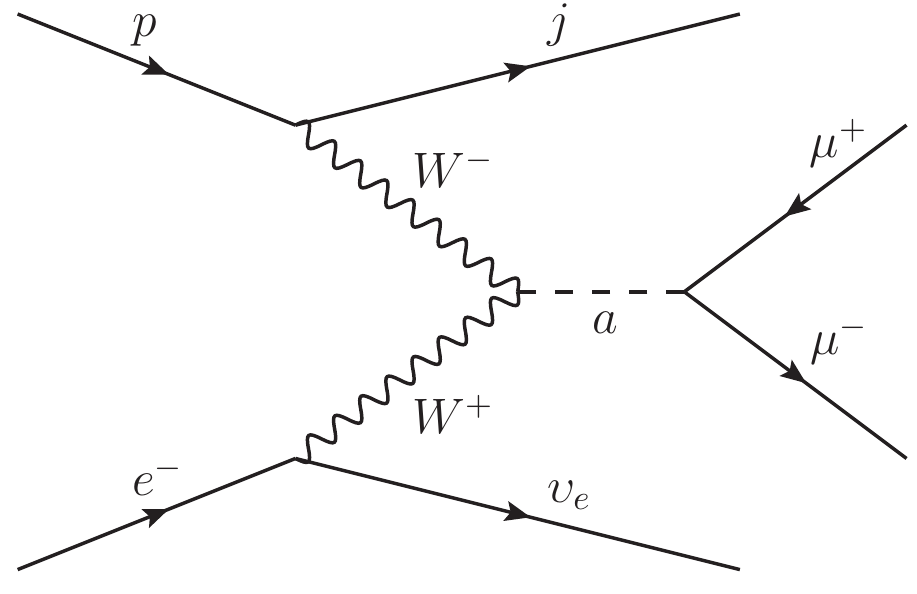}}
\hspace{0.7in}
\subfigure[]{\includegraphics [scale=0.5] {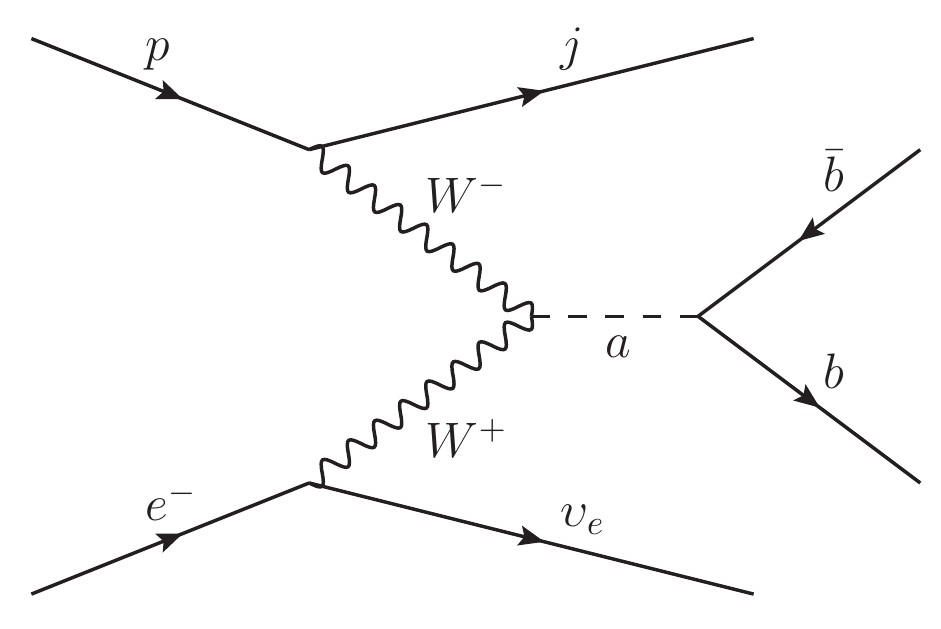}}
\caption{Illustration of the Feynman diagrams for the $W^{+}W^{-}$ fusion processes $e^{-}p\rightarrow\nu_{e}ja(a\rightarrow\mu^{+}\mu^{-})$ (a) and $e^{-}p\rightarrow\nu_{e}ja(a\rightarrow b\overline{b})$ (b) at the LHeC.}
\label{Feynmandiagrams}
\end{center}
\end{figure}

LHeC is a powerful TeV energy collider and the possibility of discovering new physics below the TeV scale could be addressed by it. Owing to good capabilities of the LHeC detector in identifying muons and performing b-tagging, the decay channels $a\rightarrow\mu^{+}\mu^{-}$ as well as $a\rightarrow b\overline{b}$ are mainly concerned, which could be complementary to the searches at the LHC~\cite{Brivio:2017ije,Craig:2018kne}. The Feynman diagrams describing the production of ALP via the $W^{+}W^{-}$ fusion process $e^{-}p\rightarrow\nu_{e}ja$ followed by $a$ decaying to muon pairs and $b\overline{b}$ pairs, respectively, are displayed in FIG.~\ref{Feynmandiagrams}. We focus on the production of ALP with mass $m_a\leq1000$ GeV at the LHeC in this paper, picking $g_{a\psi}$ equals to $1$ TeV$^{-1}$. In the following, the $60$ GeV electron beam and $7$ TeV proton beam of the LHeC are considered to obtain the center of mass energy approaching $1.3$ TeV with the integrated luminosity of $1$ ab$^{-1}$.

\section{The discovery potential for ALP of the LH\lowercase{e}C}

\subsection{Searching for ALP via $e^{-}p\rightarrow\nu_{e}ja(a\rightarrow\mu^{+}\mu^{-})$ at the LHeC  }

Our phenomenological study starts with the $W^{+}W^{-}$ fusion process $e^{-}p\rightarrow\nu_{e}ja$ followed by $a\rightarrow\mu^{+}\mu^{-}$, in which the ALP with mass from 5 GeV to 1000 GeV is taken into account. This channel provides final states that include muon pairs, a light jet $j=(u$, $d$, $c$, or $s)$ and missing energy $\slashed E$. The corresponding SM background is $e^{-}p\rightarrow\nu_{e}j\mu^{+}\mu^{-}$. A Monte Carlo (MC) simulation is performed to explore the potential of detecting ALP at the LHeC. All the signal and background events that are going to be discussed in this and the next subsection will be generated in \verb"MadGraph5_aMC@NLO"~\cite{Alwall:2014hca} with basic cuts, which require the leptons (electrons and muons included) with transverse momentum $p_{T}^{l}>10$ GeV and the jets (light flavor jets and b-jets included) with transverse momentum $p_{T}^{j}>20$ GeV. The absolute values of the leptons pseudorapidity $\eta_{l}$ and the jets pseudorapidity $\eta_{j}$ need to be less than $2.5$ and $5$, respectively. The angular separation requirements are $\Delta R _{ll}>0.4$, $\Delta R _{jj}>0.4$ as well as $\Delta R _{lj}>0.4$ for leptons and jets, in which $\Delta$$R$ is defined as $\sqrt{(\Delta \phi)^{2}+(\Delta \eta)^{2}}$. The \verb"PYTHIA8" program~\cite{Sjostrand:2014zea} is implemented for showering and hadronization. We use \verb"DELPHES"~\cite{deFavereau:2013fsa} for the fast simulation of the LHeC detector and \verb"MadAnalysis5"~\cite{Conte:2012fm,Conte:2014zja,Conte:2018vmg} for the analysis of the resulting output.

\begin{figure}[H]
\begin{center}
\subfigure[]{\includegraphics [scale=0.35] {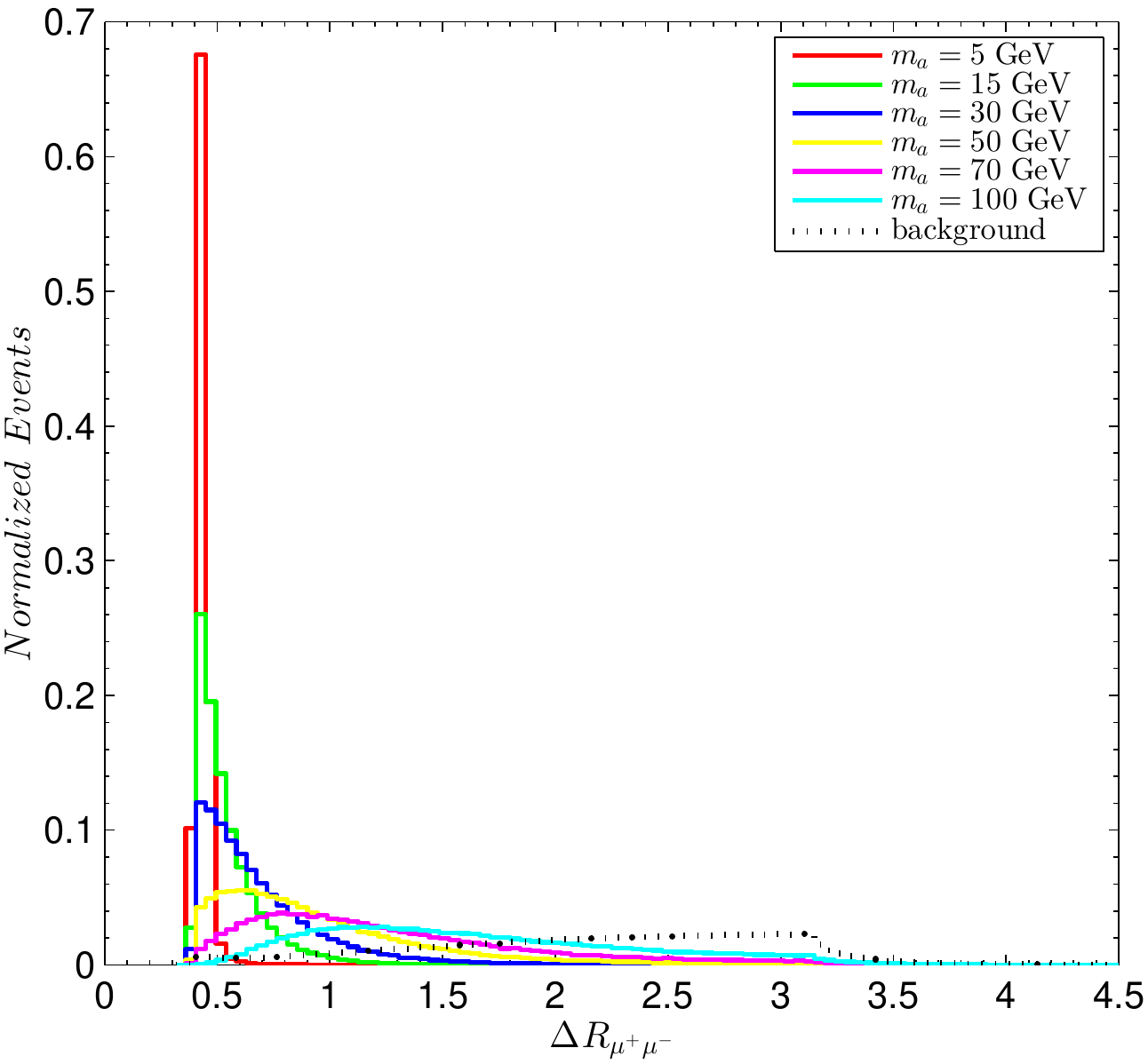}}
\hspace{0.2in}
\subfigure[]{\includegraphics [scale=0.35] {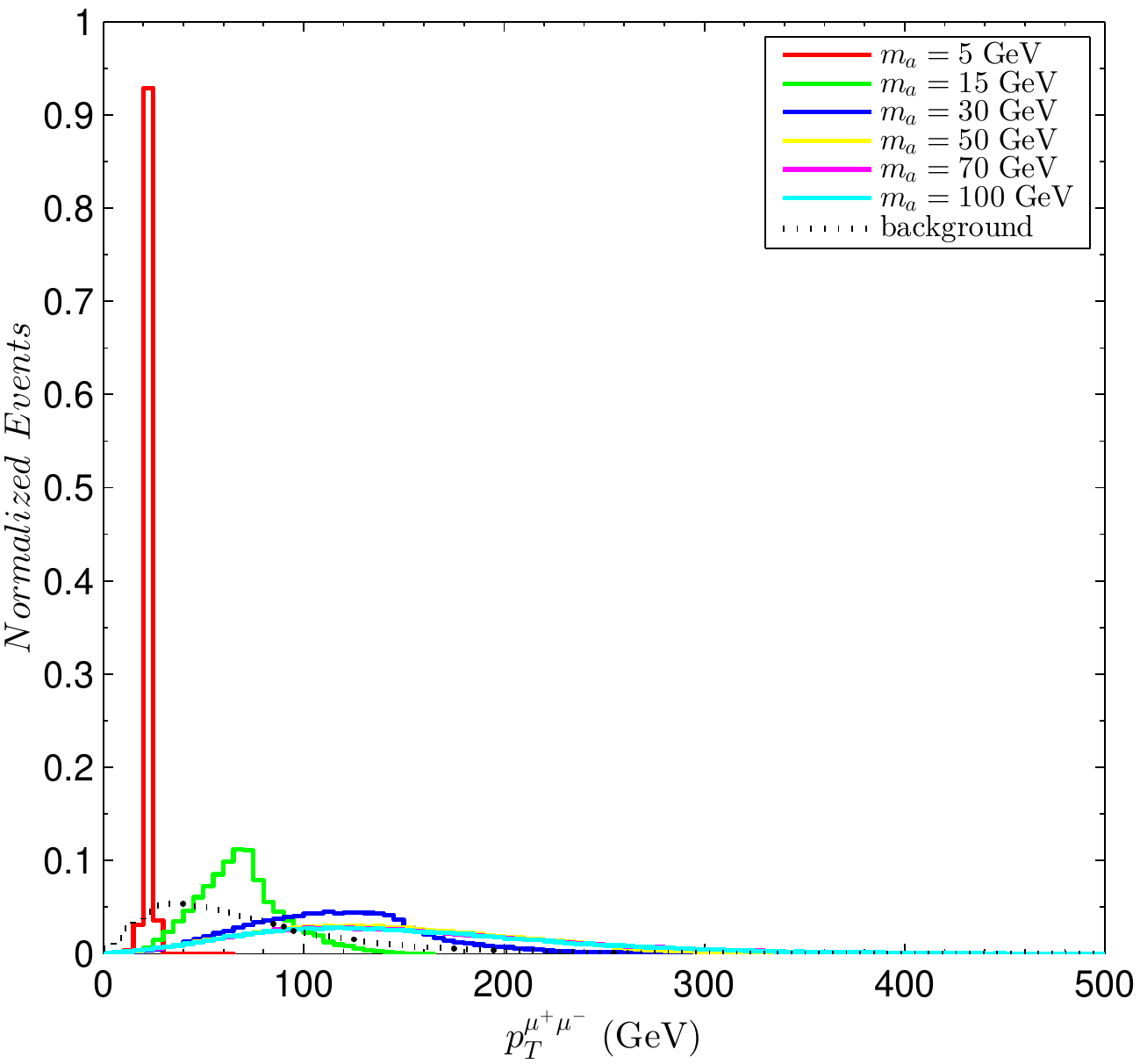}}
\caption{{The normalized distributions of the observables $\Delta R _{\mu^{+}\mu^{-}}$ (a) and $p_{T}^{\mu^{+}\mu^{-}}$ (b) in signal and background events for various ALP masses at the 1.3 TeV LHeC with $\mathcal{L}=$ $1$ ab$^{-1}$.}}
\label{distribution1}
\end{center}
\end{figure}

The cross section of the background is larger than that of our signal in the parameter region considered in this paper. The cross sections of the signal process with different $g_{aWW}$ at the $1.3$ TeV LHeC increase with the coupling coefficient $g_{aWW}$ and peak near the ALP mass of 100 GeV, therefore, we take such mass as a breakpoint to divide the mass range of ALP considered into two intervals for the study. For ALP in the mass range of $5$ GeV to $100$ GeV, observables of the angular separation between the pair of muons $\Delta R _{\mu^{+}\mu^{-}}$ in the final states and the transverse momentum of reconstructed ALP $p_{T}^{\mu^{+}\mu^{-}}$ are chosen. We present in FIG.~\ref{distribution1} the distributions of $\Delta R _{\mu^{+}\mu^{-}}$ and $p_{T}^{\mu^{+}\mu^{-}}$ for the signal and background events with typical points of $m_a = 5$, $15$, $30$, $50$, $70$, $100$ GeV. Two muons from the light ALP decay are much closer, whereas the angular separation between them tends to have a wide distribution for the background events.

\begin{figure}[H]
\begin{center}
\subfigure[]{\includegraphics [scale=0.35] {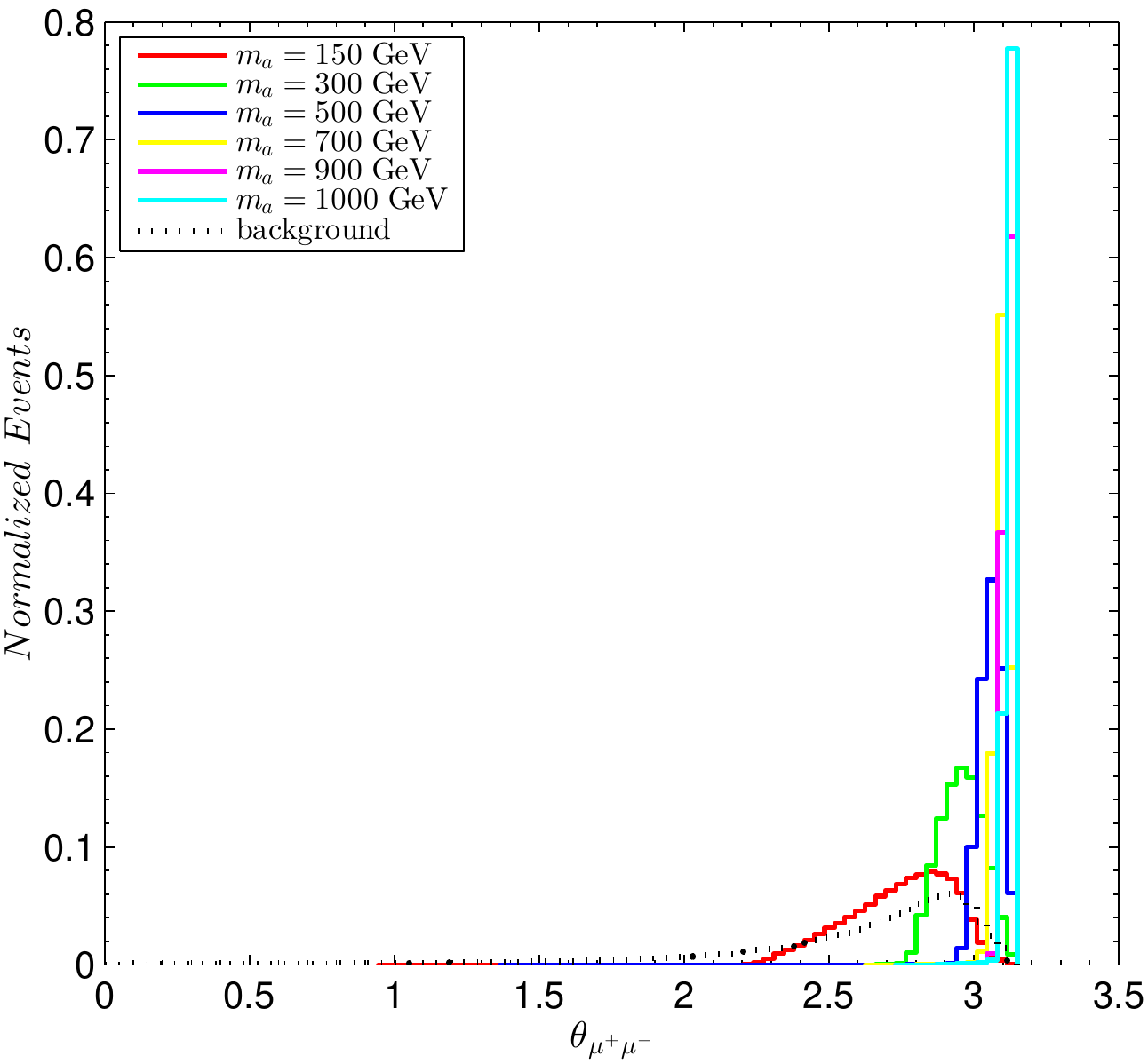}}
\hspace{0.2in}
\subfigure[]{\includegraphics [scale=0.35] {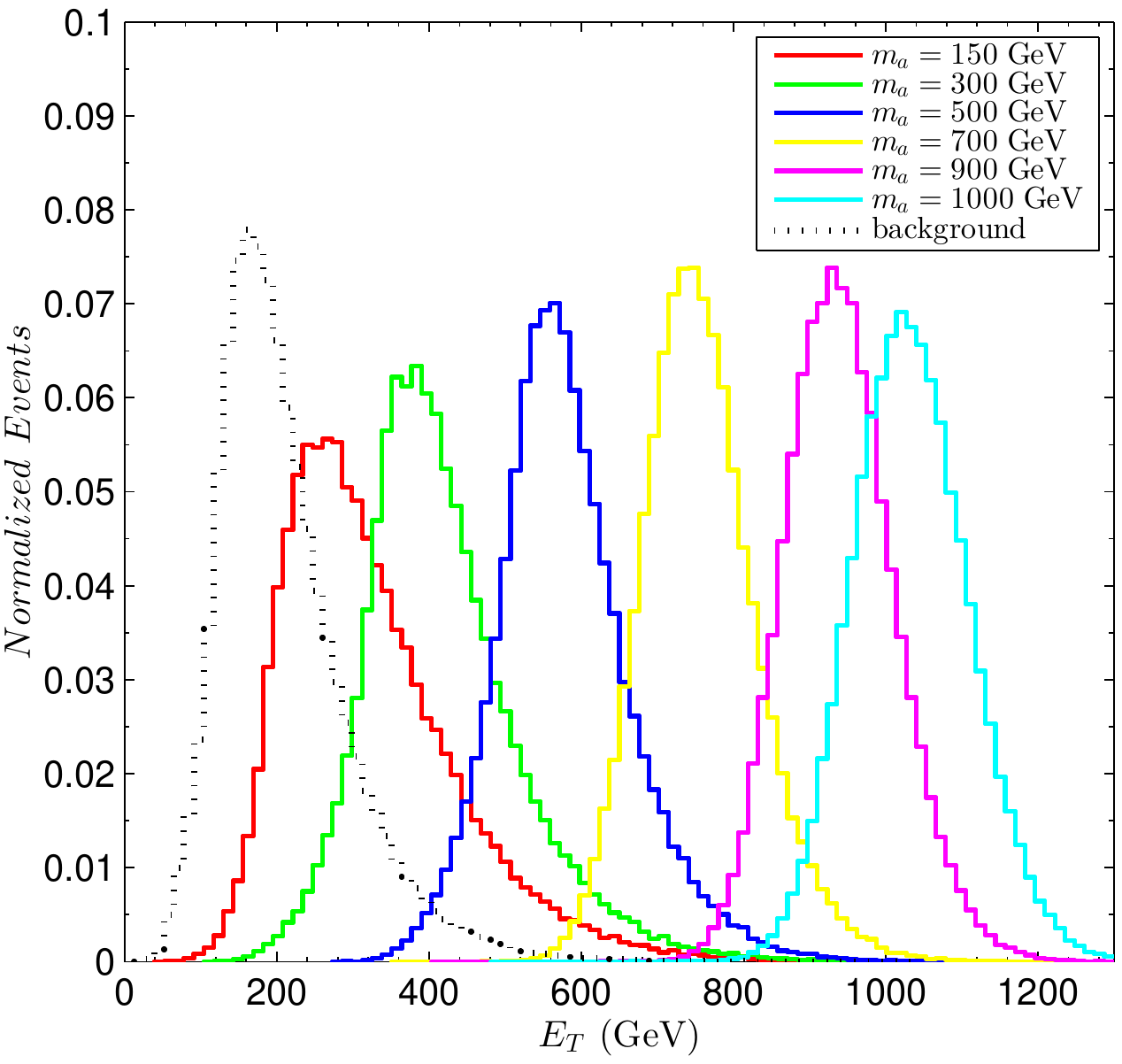}}
\hspace{0.2in}
\subfigure[]{\includegraphics [scale=0.35] {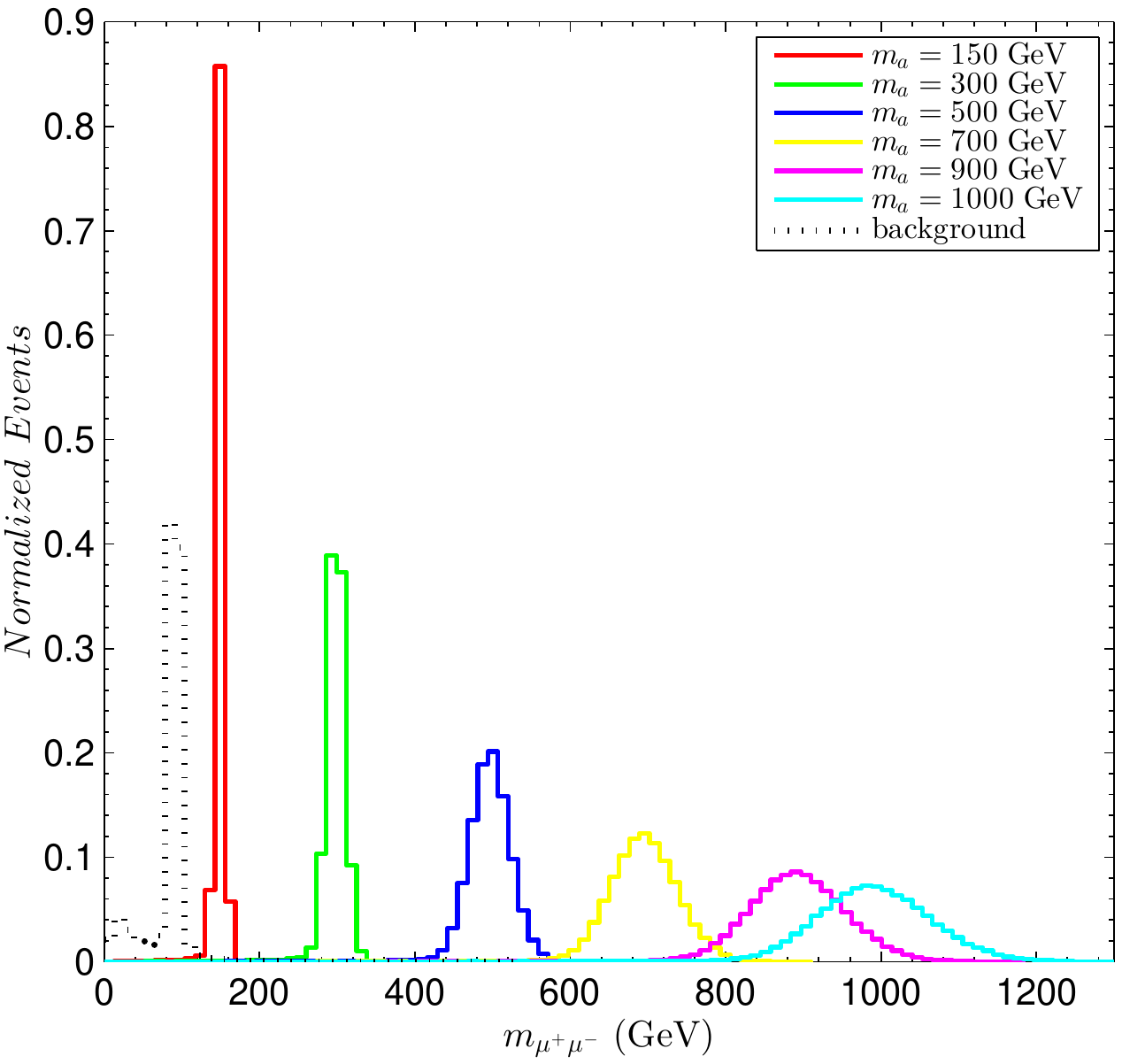}}
\caption{The normalized distributions of the observables $\theta_{\mu^{+}\mu^{-}}$ (a), $E_T$ (b) and $m_{\mu^{+}\mu^{-}}$ (c) in signal and background events for various ALP masses at the 1.3 TeV LHeC with $\mathcal{L}=$ $1$ ab$^{-1}$.}
\label{distribution2}
\end{center}
\end{figure}

Variables of the angel between reconstructed ALP and the beam axis $\theta_{\mu^{+}\mu^{-}}$, the total transverse energy of the final states $E_T$ and the invariant mass of muon pairs $m_{\mu^{+}\mu^{-}}$ are taken to be analyzed when the ALP mass falls in the region of $100$ GeV to $1000$ GeV, meanwhile the normalized distributions of them are given in FIG.~\ref{distribution2} based on six benchmark points as $m_a = 150$, $300$, $500$, $700$, $900$, $1000$ GeV. It can be seen that the signal and background  can be well distinguished in the invariant mass $m_{\mu^{+}\mu^{-}}$ distribution. The SM background mainly comes from $e^{-}p\rightarrow\nu_{e}jZ$ followed by $Z$ decaying into muon pairs. The signal events have peaks around the ALP mass, which become wider as the ALP mass increases.

According to the information of these kinematic distributions, improved cuts presented in TABLE~\ref{table:mumucut} are further imposed for separating the signal events from the background events, where the particle numbers in the final states are subject to the conditions of $N_{\mu^{+}}\geq1$, $N_{\mu^{-}}\geq1$ and $N_j\geq1$. In TABLE~\ref{tab1} and TABLE~\ref{tab2}, we show the cross sections of the signal and background at the LHeC with $\sqrt{s}=1.3$ TeV after taking the step-by-step cuts for few ALP mass benchmark points and the specific parameter $g_{aWW}=1$ TeV$^{-1}$. It can be seen that the background is effectively suppressed. We further calculate the statistical significance $SS=S/\sqrt{S + B}$ for the luminosity of $1$ ab$^{-1}$, where $S$ and $B$ are the number of events for the signal and background, respectively. Large $SS$ values can be attained in a broad region of the parameter space, as illustrated in TABLE~\ref{tab1} and TABLE~\ref{tab2}. The statistical significance can reach $7.015$ ($1.581$) for $m_a=15$ GeV ($700$ GeV).

\begin{table}[H]
\begin{center}
\setlength{\tabcolsep}{1.5mm}{
\caption{The improved cuts on signal and background for $5$ GeV $\leq$ $m_a\leq1000$ GeV.}
\label{table:mumucut}
\begin{tabular}
[c]{c|c c c c c}\hline \hline
\multirow{2}{*}{Cuts}    & \multicolumn{2}{c}{ Mass }   \\
\cline{2-3}
	       &~~~~~~~$5$ GeV $\leq$ $m_a\leq100$ GeV~~~~~~~      &~~~~~~~$100$ GeV $<$ $m_a \leq1000$ GeV~~~~~~~      \\ \hline
	Cut 1      &  $N_{\mu^{+}}\geq1$, $N_{\mu^{-}}\geq1$, $N_j\geq1$  & $N_{\mu^{+}}\geq1$, $N_{\mu^{-}}\geq1$, $N_j\geq1$ \\
	Cut 2         &  $\Delta R _{\mu^{+}\mu^{-}}<1.8$  & $\theta_{\mu^{+}\mu^{-}}>2.1$   \\

	Cut 3         &  $p_{T}^{\mu^{+}\mu^{-}}>20$ GeV  & $E_T>250$ GeV    \\

	Cut 4         &  $-$  & $m_{\mu^{+}\mu^{-}}>100$ GeV    \\ \hline \hline

\end{tabular}}
\end{center}
\end{table}

\begin{table}[H]\tiny
	\centering{
\caption{The cross sections of the $W^{+}W^{-}$ fusion process $e^{-}p\rightarrow\nu_{e}ja(a\rightarrow\mu^{+}\mu^{-})$ and the background $e^{-}p\rightarrow\nu_{e}j\mu^{+}\mu^{-}$ after the improved cuts applied for $g_{aWW}=1$ TeV$^{-1}$ at the 1.3 TeV LHeC with benchmark points.$~$\label{tab1}}
		\newcolumntype{C}[1]{>{\centering\let\newline\\\arraybackslash\hspace{50pt}}m{#1}}
		\begin{tabular}{m{1.5cm}<{\centering}|m{2cm}<{\centering} m{2cm}<{\centering} m{2cm}<{\centering}  m{2cm}<{\centering} m{2cm}<{\centering} m{2cm}<{\centering}}
			\hline \hline
      \multirow{2}{*}{Cuts} & \multicolumn{6}{c}{cross sections for signal (background) [pb]}\\
     \cline{2-7}
     & $m_a=5$ GeV  & $m_a=15$ GeV  & $m_a=30$ GeV  & $m_a=50$ GeV & $m_a=70$ GeV & $m_a=100$ GeV  \\ \hline
     Basic Cuts  & \makecell{$1.7442\times10^{-6}$\\$(0.0142)$} & \makecell{$5.2885\times10^{-4}$\\$(0.0142)$} &\makecell{$2.0437\times10^{-3}$\\$(0.0142)$} &\makecell{$2.8861\times10^{-3}$\\$(0.0142)$} & \makecell{$2.9864\times10^{-3}$\\$(0.0142)$} & \makecell{$2.7399\times10^{-3}$\\$(0.0142)$}
        \\
     Cut 1  & \makecell{$1.4435\times10^{-6}$\\$(0.0128)$} & \makecell{$4.6473\times10^{-4}$\\$(0.0128)$} &\makecell{$1.8361\times10^{-3}$\\$(0.0128)$} &\makecell{$2.6165\times10^{-3}$\\$(0.0128)$} & \makecell{$2.7072\times10^{-3}$\\$(0.0128)$} & \makecell{$2.4812\times10^{-3}$\\$(0.0128)$}
       \\
     Cut 2  & \makecell{$1.4435\times10^{-6}$\\$(3.9212\times10^{-3})$} & \makecell{$4.6463\times10^{-4}$\\$(3.9212\times10^{-3})$} &\makecell{$1.8156\times10^{-3}$\\$(3.9212\times10^{-3})$} &\makecell{$2.4350\times10^{-3}$\\$(3.9212\times10^{-3})$} & \makecell{$2.2205\times10^{-3}$\\$(3.9212\times10^{-3})$} & \makecell{$1.5666\times10^{-3}$\\$(3.9212\times10^{-3})$}
       \\
     Cut 3  & \makecell{$1.3934\times10^{-6}$\\$(3.8963\times10^{-3})$} & \makecell{$4.6375\times10^{-4}$\\$(3.8963\times10^{-3})$} &\makecell{$1.8155\times10^{-3}$\\$(3.8963\times10^{-3})$} &\makecell{$2.4349\times10^{-3}$\\$(3.8963\times10^{-3})$} & \makecell{$2.2205\times10^{-3}$\\$(3.8963\times10^{-3})$} & \makecell{$1.5666\times10^{-3}$\\$(3.8963\times10^{-3})$}
     \\ \hline
     $SS$  & $0.022$ & $7.015$ & $24.022$ & $30.609$ & $28.396$ & $21.196$\\ \hline \hline
	\end{tabular}}	
\end{table}

\begin{table}[H]\tiny
	\centering{
\caption{Same as TABLE~\ref{tab1} but for benchmark points $m_a = 150$, $300$, $500$, $700$, $900$, $1000$ GeV.$~$\label{tab2}}
		\newcolumntype{C}[1]{>{\centering\let\newline\\\arraybackslash\hspace{50pt}}m{#1}}
		\begin{tabular}{m{1.5cm}<{\centering}|m{2cm}<{\centering} m{2cm}<{\centering} m{2cm}<{\centering}  m{2cm}<{\centering} m{2cm}<{\centering} m{2cm}<{\centering}}
			\hline \hline
      \multirow{2}{*}{Cuts} & \multicolumn{6}{c}{cross sections for signal (background) [pb]}\\
     \cline{2-7}
     & $m_a=150$ GeV  & $m_a=300$ GeV  & $m_a=500$ GeV  & $m_a=700$ GeV & $m_a=900$ GeV & $m_a=1000$ GeV  \\ \hline
     Basic Cuts  & \makecell{$2.1509\times10^{-3}$\\$(0.0142)$} & \makecell{$8.8581\times10^{-4}$\\$(0.0142)$} &\makecell{$2.0197\times10^{-4}$\\$(0.0142)$} &\makecell{$2.7954\times10^{-5}$\\$(0.0142)$} & \makecell{$1.5617\times10^{-6}$\\$(0.0142)$} & \makecell{$1.9118\times10^{-7}$\\$(0.0142)$}
        \\
     Cut 1  & \makecell{$1.9429\times10^{-3}$\\$(0.0128)$} & \makecell{$7.9889\times10^{-4}$\\$(0.0128)$} &\makecell{$1.8218\times10^{-4}$\\$(0.0128)$} &\makecell{$2.4999\times10^{-5}$\\$(0.0128)$} & \makecell{$1.3965\times10^{-6}$\\$(0.0128)$} & \makecell{$1.7116\times10^{-7}$\\$(0.0128)$}
       \\
     Cut 2  & \makecell{$1.9417\times10^{-3}$\\$(0.0112)$} &\makecell{$7.9882\times10^{-4}$\\$(0.0112)$} &\makecell{$1.8218\times10^{-4}$\\$(0.0112)$} & \makecell{$2.4999\times10^{-5}$\\$(0.0112)$} & \makecell{$1.3965\times10^{-6}$\\$(0.0112)$}&
     \makecell{$1.7116\times10^{-7}$\\$(0.0112)$}
       \\
     Cut 3 & \makecell{$1.3790\times10^{-3}$\\$(3.2070\times10^{-3})$} &\makecell{$7.7811\times10^{-4}$\\$(3.2070\times10^{-3})$} &\makecell{$1.8218\times10^{-4}$\\$(3.2070\times10^{-3})$} & \makecell{$2.4999\times10^{-5}$\\$(3.2070\times10^{-3})$} & \makecell{$1.3965\times10^{-6}$\\$(3.2070\times10^{-3})$} &
     \makecell{$1.7116\times10^{-7}$\\$(3.2070\times10^{-3})$}
     \\
     Cut 4 & \makecell{$1.3713\times10^{-3}$\\$(2.2571\times10^{-4})$} &\makecell{$7.7691\times10^{-4}$\\$(2.2571\times10^{-4})$} &\makecell{$1.8208\times10^{-4}$\\$(2.2571\times10^{-4})$} & \makecell{$2.4989\times10^{-5}$\\$(2.2571\times10^{-4})$} & \makecell{$1.3965\times10^{-6}$\\$(2.2571\times10^{-4})$} &
     \makecell{$1.7116\times10^{-7}$\\$(2.2571\times10^{-4})$}
     \\ \hline
     $SS$  & $34.309$ & $24.526$ & $8.986$ & $1.581$ & $0.093$ & $0.011$\\ \hline \hline
	\end{tabular}}	
\end{table}

\subsection{Searching for ALP via $e^{-}p\rightarrow\nu_{e}ja(a\rightarrow b\overline{b})$ at the LHeC  }

In this subsection, the approach to the analysis performed in studying the channel of the ALP decays to $b\overline{b}$ pairs is similar to that used in the $a\rightarrow \mu^{+}\mu^{-}$ decay mode, but the ALP mass range is chosen to be $15$ GeV to $1000$ GeV. The SM backgrounds are dominantly from  $\nu_{e}jb\overline{b}$ and $\nu_{e}jjj$ for the signal process $e^{-}p\rightarrow\nu_{e}ja(a\rightarrow b\overline{b})$, in which the $\nu_{e}jjj$ final states are more severe than the $\nu_{e}jb\overline{b}$ final states. The cross sections of the signal are several orders of magnitude smaller than the corresponding backgrounds after the basic cuts applied.

\begin{figure}[H]
\begin{center}
\subfigure[]{\includegraphics [scale=0.35] {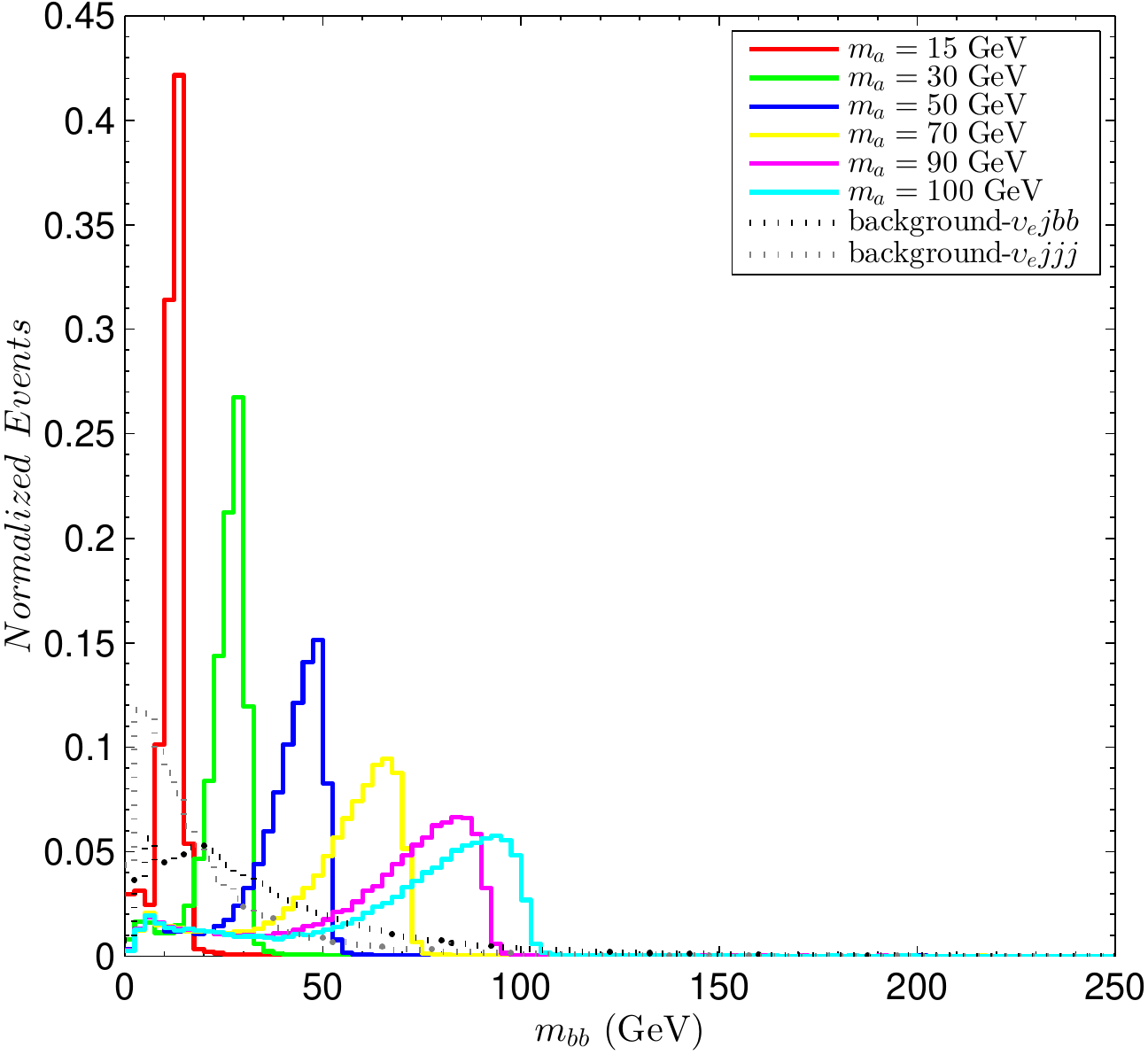}}
\hspace{0.2in}
\subfigure[]{\includegraphics [scale=0.35] {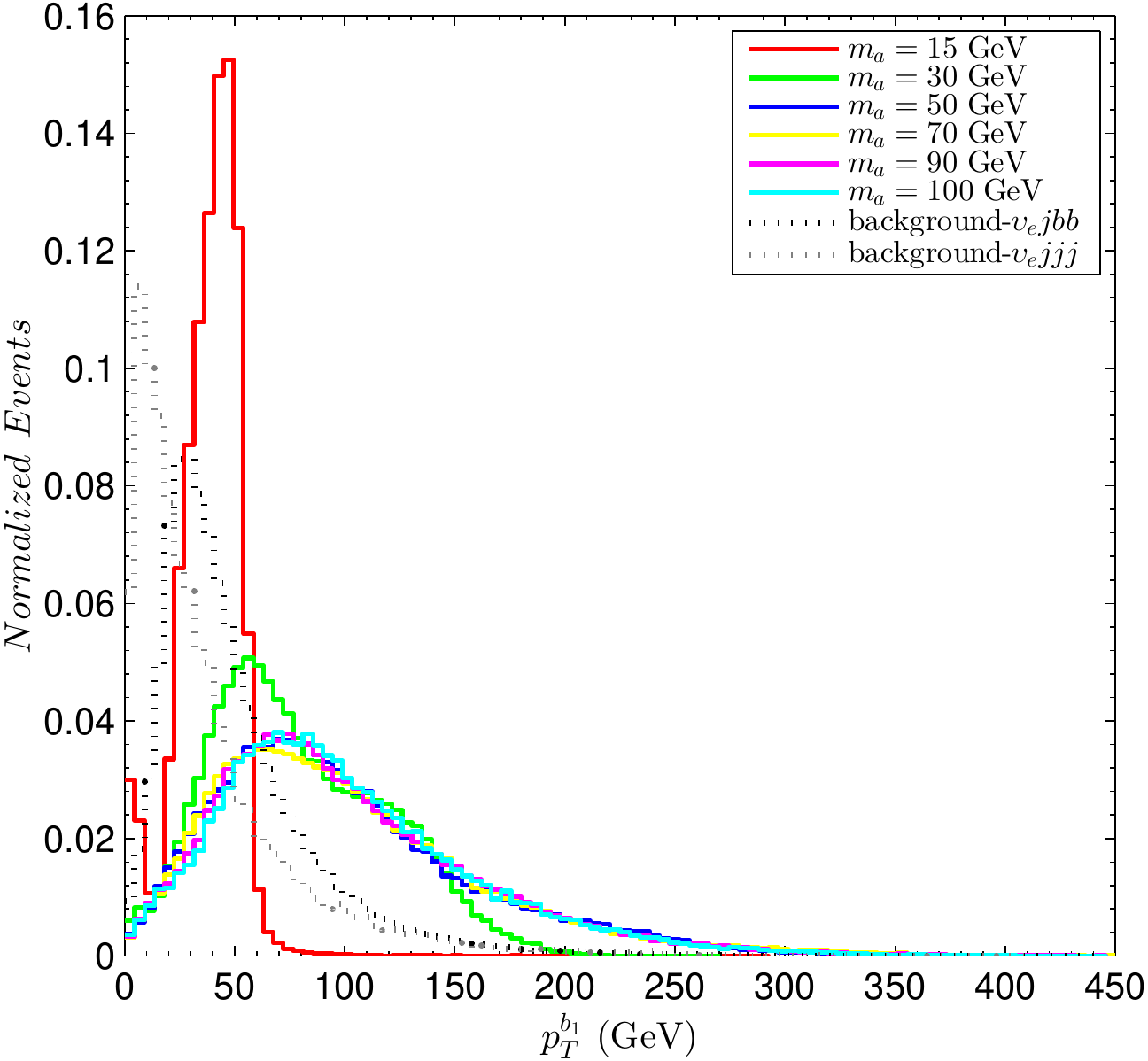}}
\hspace{0.8in}
\subfigure[]{\includegraphics [scale=0.35] {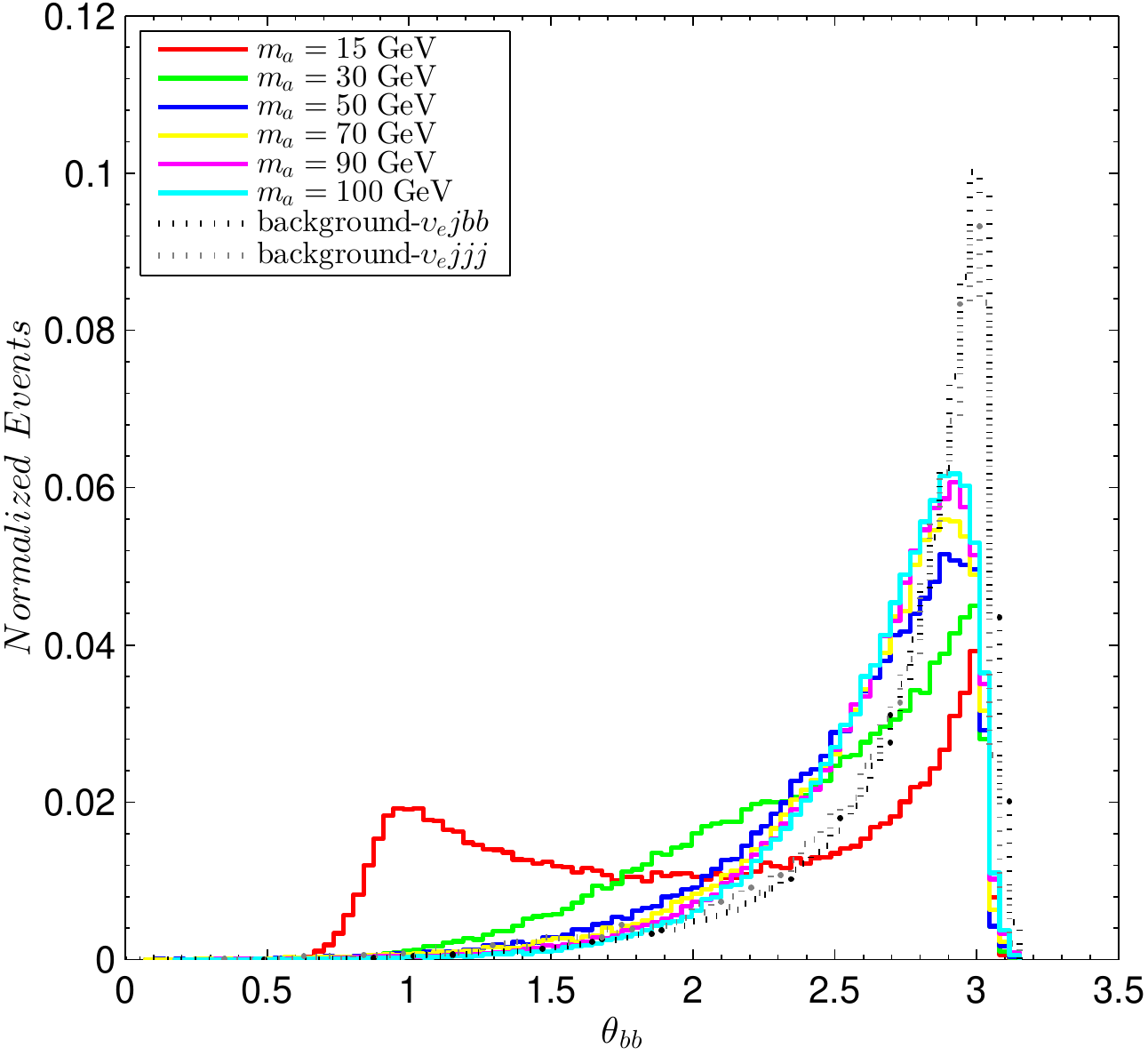}}
\hspace{0.2in}
\subfigure[]{\includegraphics [scale=0.35] {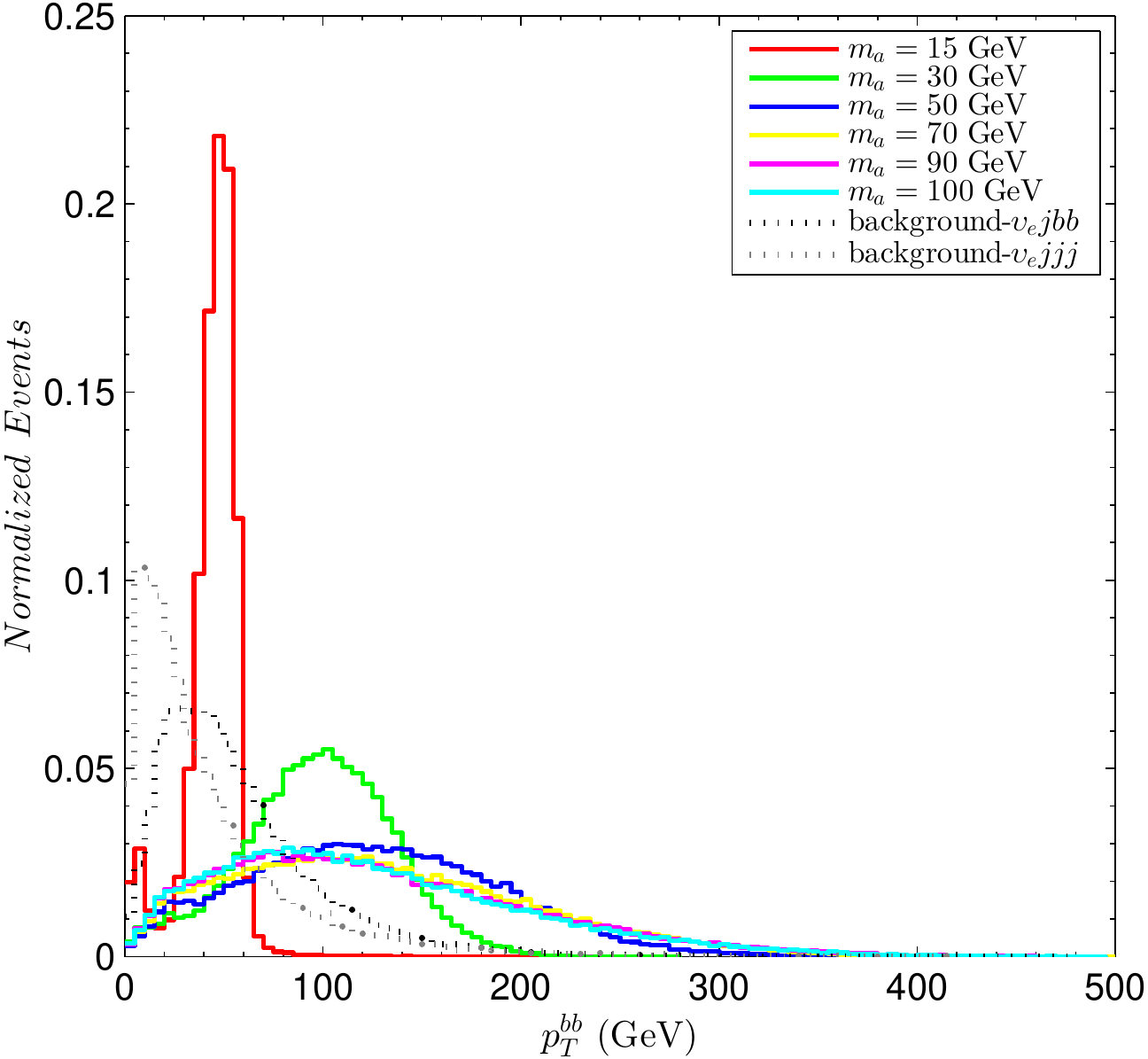}}
\caption{The normalized distributions of the observables $m_{bb}$ (a), $p_{T}^{b_{1}}$ (b), $\theta_{bb}$ (c) and $p_{T}^{bb}$ (d) in signal and background events for various ALP masses at the 1.3 TeV LHeC with $\mathcal{L}=$ $1$ ab$^{-1}$.}
\label{distribution3}
\end{center}
\end{figure}

\begin{figure}[H]
\begin{center}
\subfigure[]{\includegraphics [scale=0.35] {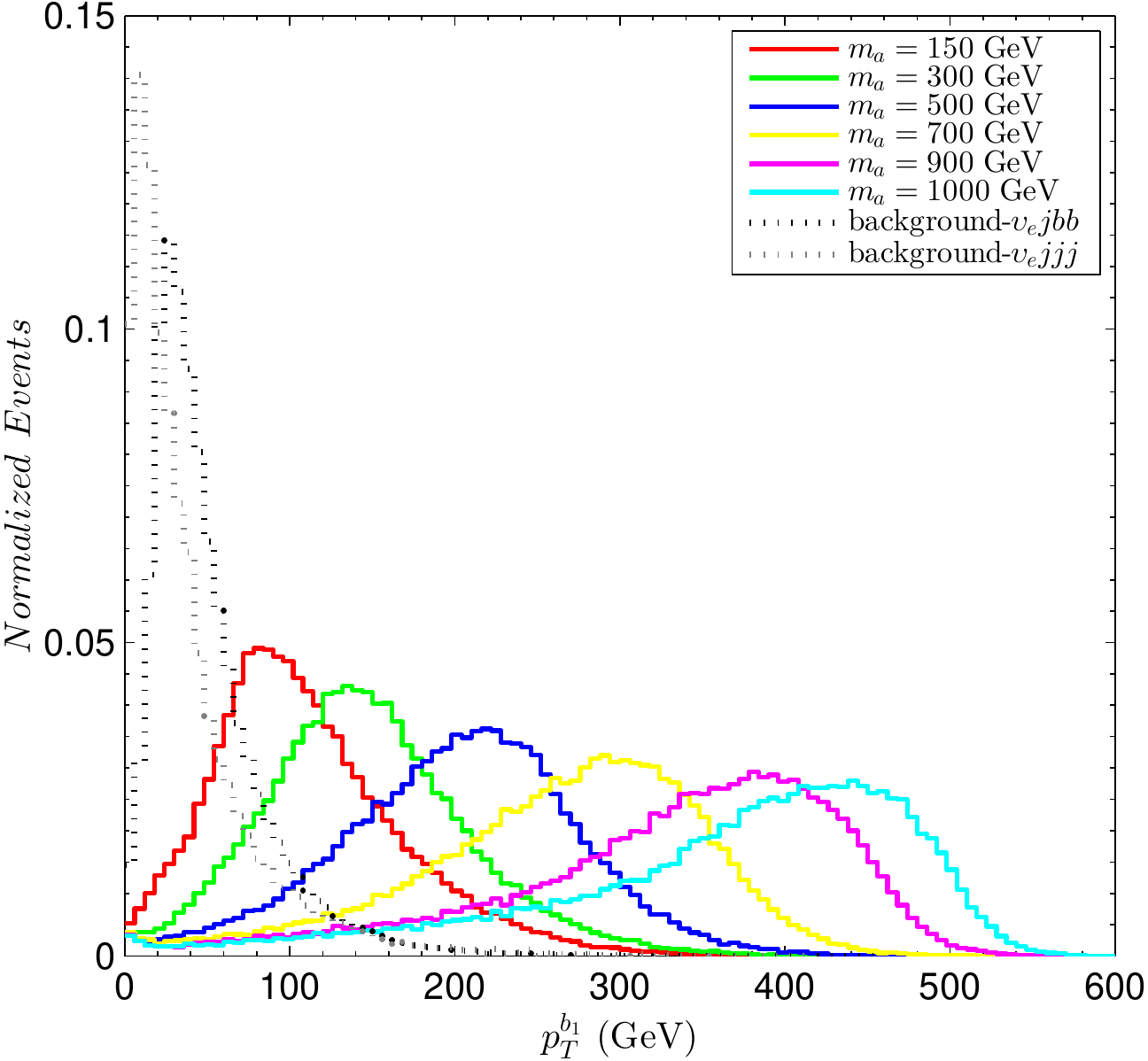}}
\hspace{0.2in}
\subfigure[]{\includegraphics [scale=0.35] {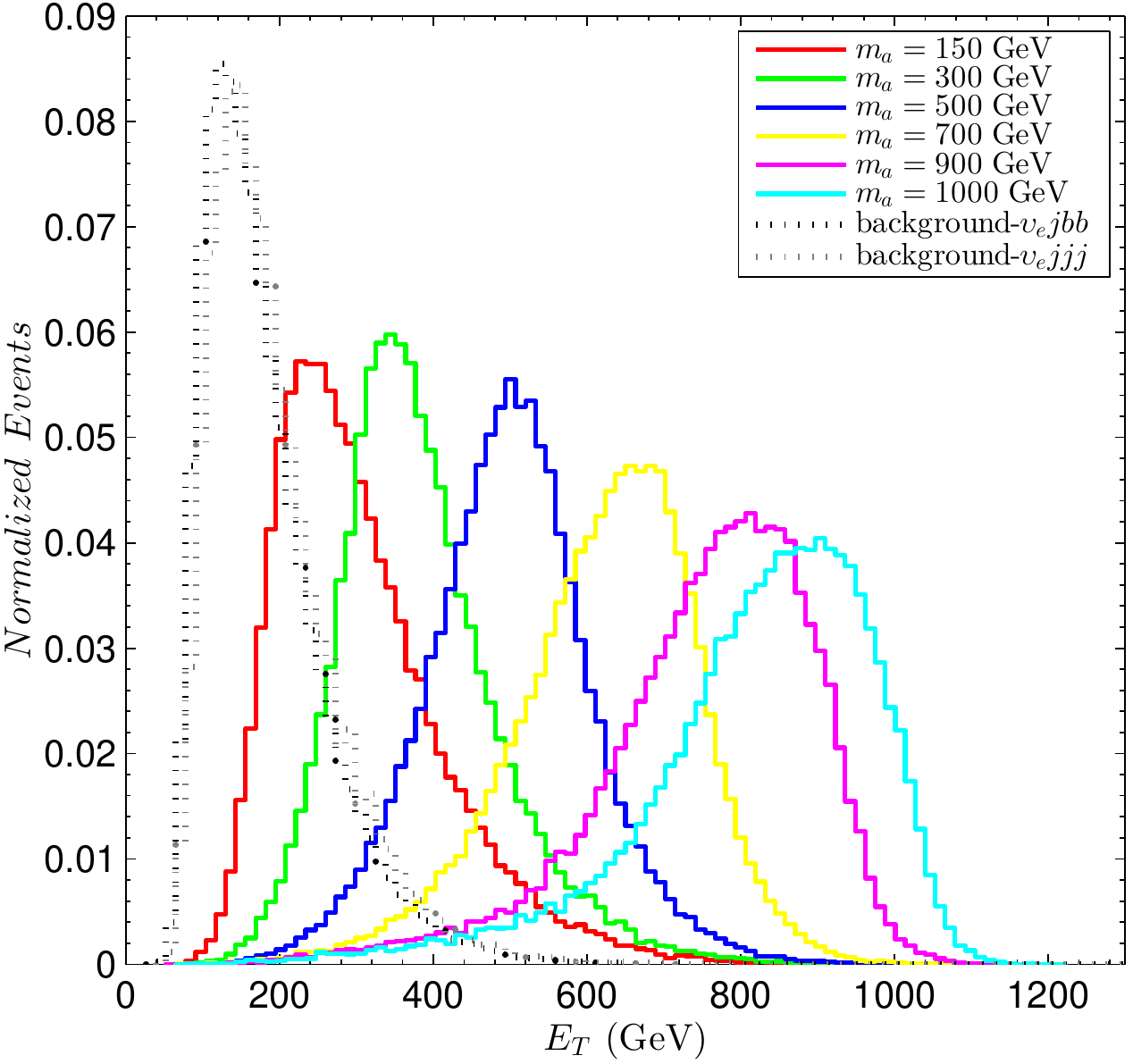}}
\hspace{0.2in}
\subfigure[]{\includegraphics [scale=0.35] {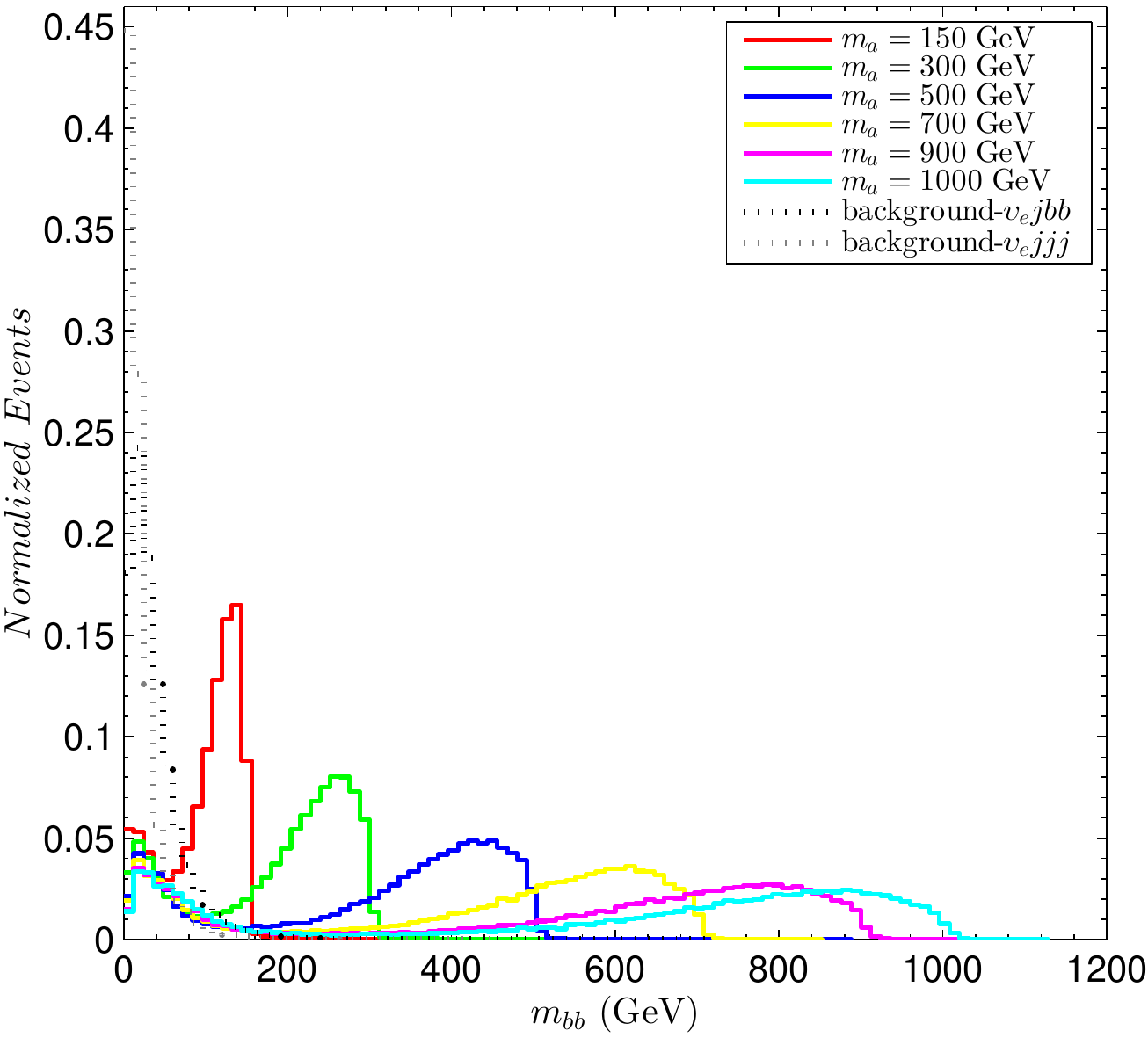}}
\caption{The normalized distributions of the observables $p_{T}^{b_{1}}$ (a), $E_T$ (b) and $m_{bb}$ (c) in signal and background events for various ALP masses at the 1.3 TeV LHeC with $\mathcal{L}=$ $1$ ab$^{-1}$.}
\label{distribution4}
\end{center}
\end{figure}

Even though the signal could easily be overwhelmed by the enormous backgrounds, there are many kinematic differences between them that can be utilized to distinguish the signal from the backgrounds. The invariant mass of $bb$ pairs $m_{bb}$, the transverse momentum of the hardest b-jet $p_{T}^{b_{1}}$, the angel between reconstructed ALP and the beam axis $\theta_{bb}$ as well as the transverse momentum of reconstructed ALP $p_{T}^{bb}$ are taken as the variables for analysis in the mass interval of $15$ GeV to $100$ GeV for ALP. The normalized distributions of these kinematic variables are exhibited in FIG.~\ref{distribution3}, where some ALP mass benchmark points $m_a = 15$, $30$, $50$, $70$, $90$, $100$ GeV are picked as examples. The transverse momentum of the reconstructed $bb$ pairs is small for the background events, while this is not the case for the signal. Some selected kinematic distributions, namely $p_{T}^{b_{1}}$, the total transverse energy of the final states $E_T$ and $m_{bb}$, for the signal and background events with six mass points in $100$ GeV $<$ $m_a\leq1000$ GeV are shown in FIG.~\ref{distribution4}. We can see that the $p_{T}^{b_{1}}$ peaks of the signal events are larger than those of the SM backgrounds.

In order to trigger the signal events, different optimized kinematical cuts are applied to reduce backgrounds and improve the statistical significance, as listed in Table~\ref{table:bbcut}. The estimation of cross sections after applying the above selection cuts for the signal and potential background processes are given in TABLE~\ref{tab3} and TABLE~\ref{tab4}, in which the $\nu_{e}jb\overline{b}$ background and the $\nu_{e}jjj$ background are labelled as ``background1" and ``background2", respectively. The $SS$ values obtained with the selection strategy are summarized in the last row of TABLE~\ref{tab3} and TABLE~\ref{tab4}. A tagging efficiency of $75\%$ for b-jets and a mistagging rate of $5\%$ for c-jets as well as $0.1\%$ for other light flavor jets are assumed. The significance of $3.468$ can be obtained when we take $m_a$ as $30$ GeV and the integrated luminosity of $1$ ab$^{-1}$. There are small values of $SS$ when the ALP mass approaching $1000$ GeV.

\begin{table}[H]
\begin{center}
\caption{The improved cuts on signal and background for $15$ GeV $\leq$ $m_a\leq1000$ GeV.}
\label{table:bbcut}
\begin{tabular}
[c]{c|c c c c c}\hline \hline
\multirow{2}{*}{Cuts}    & \multicolumn{2}{c}{ Mass }   \\
\cline{2-3}
	       &~~~~~~~$15$ GeV $\leq$ $m_a\leq100$ GeV~~~~~~~      &~~~~~~~$100$ GeV $<$ $m_a\leq1000$ GeV~~~~~~~      \\ \hline
	Cut 1      &  $N_b\geq2$, $N_j\geq1$  & $N_b\geq2$, $N_j\geq1$ \\

	Cut 2         &  $m_{bb}>10$ GeV   & $p_{T}^{b_{1}}>85$ GeV  \\

	Cut 3         &  $p_{T}^{b_{1}}>30$ GeV  & $E_T>200$ GeV    \\

    Cut 4         &  $\theta_{bb}<2.9$  & $m_{bb}>100$ GeV    \\

	Cut 5         &  $p_{T}^{bb}>40$ GeV  & $-$   \\ \hline \hline

\end{tabular}
\end{center}
\end{table}

\begin{table}[H]\tiny
	\centering{
\caption{The cross sections of the $W^{+}W^{-}$ fusion process $e^{-}p\rightarrow\nu_{e}ja(a\rightarrow b\overline{b})$ and the background processes $e^{-}p\rightarrow\nu_{e}jb\overline{b}$ and $e^{-}p\rightarrow\nu_{e}jjj$ after the improved cuts applied for $g_{aWW}=1$ TeV$^{-1}$ at the 1.3 TeV LHeC with benchmark points.$~$\label{tab3}}
		\newcolumntype{C}[1]{>{\centering\let\newline\\\arraybackslash\hspace{50pt}}m{#1}}
		\begin{tabular}{m{1.5cm}<{\centering}|m{2cm}<{\centering} m{2cm}<{\centering} m{2cm}<{\centering}  m{2cm}<{\centering} m{2cm}<{\centering} m{2cm}<{\centering}}
			\hline \hline
      \multirow{2}{*}{Cuts} & \multicolumn{6}{c}{cross sections for signal (background1, background2) [pb]}\\
     \cline{2-7}
     & $m_a=15$ GeV  & $m_a=30$ GeV  & $m_a=50$ GeV & $m_a=70$ GeV &  $m_a=90$ GeV & $m_a=100$ GeV  \\ \hline
     Basic Cuts  & \makecell{$6.1000\times10^{-5}$\\$(0.0490, 6.2968)$} & \makecell{$1.5831\times10^{-3}$\\$(0.0490, 6.2968)$} &\makecell{$2.8332\times10^{-3}$\\$(0.0490, 6.2968)$} &\makecell{$3.0975\times10^{-3}$\\$(0.0490, 6.2968)$} & \makecell{$3.1289\times10^{-3}$\\$(0.0490, 6.2968)$} & \makecell{$3.1248\times10^{-3}$\\$(0.0490, 6.2968)$}
        \\
     Cut 1  & \makecell{$2.8250\times10^{-5}$\\$(0.0271, 0.0922)$} & \makecell{$8.1835\times10^{-4}$\\$(0.0271, 0.0922)$} &\makecell{$1.5797\times10^{-3}$\\$(0.0271, 0.0922)$} &\makecell{$1.8036\times10^{-3}$\\$(0.0271, 0.0922)$} & \makecell{$1.8604\times10^{-3}$\\$(0.0271, 0.0922)$} & \makecell{$1.8691\times10^{-3}$\\$(0.0271, 0.0922)$}
       \\
     Cut 2  &\makecell{$2.2980\times10^{-5}$\\$(0.0231, 0.0576)$}  & \makecell{$7.7615\times10^{-4}$\\$(0.0231, 0.0576)$} &\makecell{$1.4986\times10^{-3}$\\$(0.0231, 0.0576)$} &\makecell{$1.7120\times10^{-3}$\\$(0.0231, 0.0576)$} & \makecell{$1.7665\times10^{-3}$\\$(0.0231, 0.0576)$} & \makecell{$1.7751\times10^{-3}$\\$(0.0231, 0.0576)$}
       \\
     Cut 3  & \makecell{$1.9960\times10^{-5}$\\$(0.0177, 0.0387)$} & \makecell{$7.4325\times10^{-4}$\\$(0.0177, 0.0387)$} &\makecell{$1.4409\times10^{-3}$\\$(0.0177, 0.0387)$} &\makecell{$1.6500\times10^{-3}$\\$(0.0177, 0.0387)$} & \makecell{$1.7096\times10^{-3}$\\$(0.0177, 0.0387)$} & \makecell{$1.7228\times10^{-3}$\\$(0.0177, 0.0387)$}
     \\
     Cut 4  & \makecell{$1.7110\times10^{-5}$\\$(9.9450\times10^{-3},$\\$0.0250)$} & \makecell{$6.1904\times10^{-4}$\\$(9.9450\times10^{-3},$\\$0.0250)$} &\makecell{$1.1660\times10^{-3}$\\$(9.9450\times10^{-3},$\\$ 0.0250)$} &\makecell{$1.3126\times10^{-3}$\\$(9.9450\times10^{-3},$\\$0.0250)$} & \makecell{$1.3244\times10^{-3}$\\$(9.9450\times10^{-3},$\\$0.0250)$} & \makecell{$1.3193\times10^{-3}$\\$(9.9450\times10^{-3},$\\$0.0250)$}
     \\
     Cut 5  & \makecell{$1.6290\times10^{-5}$\\$(8.9469\times10^{-3},$\\$0.0220)$} & \makecell{$6.1644\times10^{-4}$\\$(8.9469\times10^{-3},$\\$0.0220)$} &\makecell{$1.1534\times10^{-3}$\\$(8.9469\times10^{-3},$\\$0.0220)$} &\makecell{$1.2712\times10^{-3}$\\$(8.9469\times10^{-3},$\\$0.0220)$} & \makecell{$1.2725\times10^{-3}$\\$(8.9469\times10^{-3},$\\$0.0220)$} & \makecell{$1.2629\times10^{-3}$\\$(8.9469\times10^{-3},$\\$0.0220)$}
     \\ \hline
     $SS$  & $0.093$ & $3.468$ & $6.434$ & $7.078$ & $7.084$ & $7.032$ \\ \hline \hline
	\end{tabular}}	
\end{table}

\begin{table}[H]\tiny
	\centering{
\caption{Same as TABLE~\ref{tab3} but for benchmark points $m_a = 150$, $300$, $500$, $700$, $900$, $1000$ GeV.$~$\label{tab4}}
		\newcolumntype{C}[1]{>{\centering\let\newline\\\arraybackslash\hspace{50pt}}m{#1}}
		\begin{tabular}{m{1.5cm}<{\centering}|m{2cm}<{\centering} m{2cm}<{\centering} m{2cm}<{\centering}  m{2cm}<{\centering} m{2cm}<{\centering} m{2cm}<{\centering}}
			\hline \hline
      \multirow{2}{*}{Cuts} & \multicolumn{6}{c}{cross sections for signal (background1, background2) [pb]}\\
     \cline{2-7}
     & $m_a=150$ GeV  & $m_a=300$ GeV  & $m_a=500$ GeV  & $m_a=700$ GeV & $m_a=900$ GeV & $m_a=1000$ GeV  \\ \hline
     Basic Cuts  & \makecell{$2.8355\times10^{-3}$\\$(0.0490, 6.2968)$} & \makecell{$1.3969\times10^{-3}$\\$(0.0490, 6.2968)$} &\makecell{$3.8331\times10^{-4}$\\$(0.0490, 6.2968)$} &\makecell{$6.6391\times10^{-5}$\\$(0.0490, 6.2968)$} & \makecell{$4.9981\times10^{-6}$\\$(0.0490, 6.2968)$} & \makecell{$7.7509\times10^{-7}$\\$(0.0490, 6.2968)$}
        \\
     Cut 1  & \makecell{$1.7067\times10^{-3}$\\$(0.0271, 0.0922)$} & \makecell{$8.2443\times10^{-4}$\\$(0.0271, 0.0922)$} &\makecell{$2.1850\times10^{-4}$\\$(0.0271, 0.0922)$} &\makecell{$3.6355\times10^{-5}$\\$(0.0271, 0.0922)$} & \makecell{$2.5790\times10^{-6}$\\$(0.0271, 0.0922)$} & \makecell{$3.8905\times10^{-7}$\\$(0.0271, 0.0922)$}
       \\
     Cut 2  & \makecell{$1.0934\times10^{-3}$\\$(3.5966\times10^{-3},$\\$9.6278\times10^{-3})$} &\makecell{$6.9825\times10^{-4}$\\$(3.5966\times10^{-3},$\\$9.6278\times10^{-3})$} &\makecell{$2.0489\times10^{-4}$\\$(3.5966\times10^{-3},$\\$9.6278\times10^{-3})$} & \makecell{$3.4825\times10^{-5}$\\$(3.5966\times10^{-3},$\\$9.6278\times10^{-3})$} & \makecell{$2.4991\times10^{-6}$\\$(3.5966\times10^{-3},$\\$9.6278\times10^{-3})$}&
     \makecell{$3.7804\times10^{-7}$\\$(3.5966\times10^{-3},$\\$9.6278\times10^{-3})$}
       \\
     Cut 3 & \makecell{$1.0447\times10^{-3}$\\$(2.7793\times10^{-3},$\\$7.7577\times10^{-3})$} &\makecell{$6.9545\times10^{-4}$\\$(2.7793\times10^{-3},$\\$7.7577\times10^{-3})$} &\makecell{$2.0480\times10^{-4}$\\$(2.7793\times10^{-3},$\\$7.7577\times10^{-3})$} & \makecell{$3.4825\times10^{-5}$\\$(2.7793\times10^{-3},$\\$7.7577\times10^{-3})$} & \makecell{$2.4991\times10^{-6}$\\$(2.7793\times10^{-3},$\\$7.7577\times10^{-3})$} &
     \makecell{$3.7804\times10^{-7}$\\$(2.7793\times10^{-3},$\\$7.7577\times10^{-3})$}
     \\
     Cut 4 & \makecell{$7.8474\times10^{-4}$\\$(8.1555\times10^{-4},$\\$8.8780\times10^{-4})$} &\makecell{$5.9819\times10^{-4}$\\$(8.1555\times10^{-4},$\\$8.8780\times10^{-4})$} &\makecell{$1.7398\times10^{-4}$\\$(8.1555\times10^{-4},$\\$8.8780\times10^{-4})$} & \makecell{$2.9176\times10^{-5}$\\$(8.1555\times10^{-4},$\\$8.8780\times10^{-4})$} & \makecell{$2.0992\times10^{-6}$\\$(8.1555\times10^{-4},$\\$8.8780\times10^{-4})$} &
     \makecell{$3.1304\times10^{-7}$\\$(8.1555\times10^{-4},$\\$8.8780\times10^{-4})$}
     \\ \hline
     $SS$  & $15.728$ & $12.461$ & $4.012$ & $0.701$ & $0.050$ & $0.008$\\ \hline \hline
	\end{tabular}}	
\end{table}

\begin{figure}[H]
\begin{center}
\centering\includegraphics [scale=0.41] {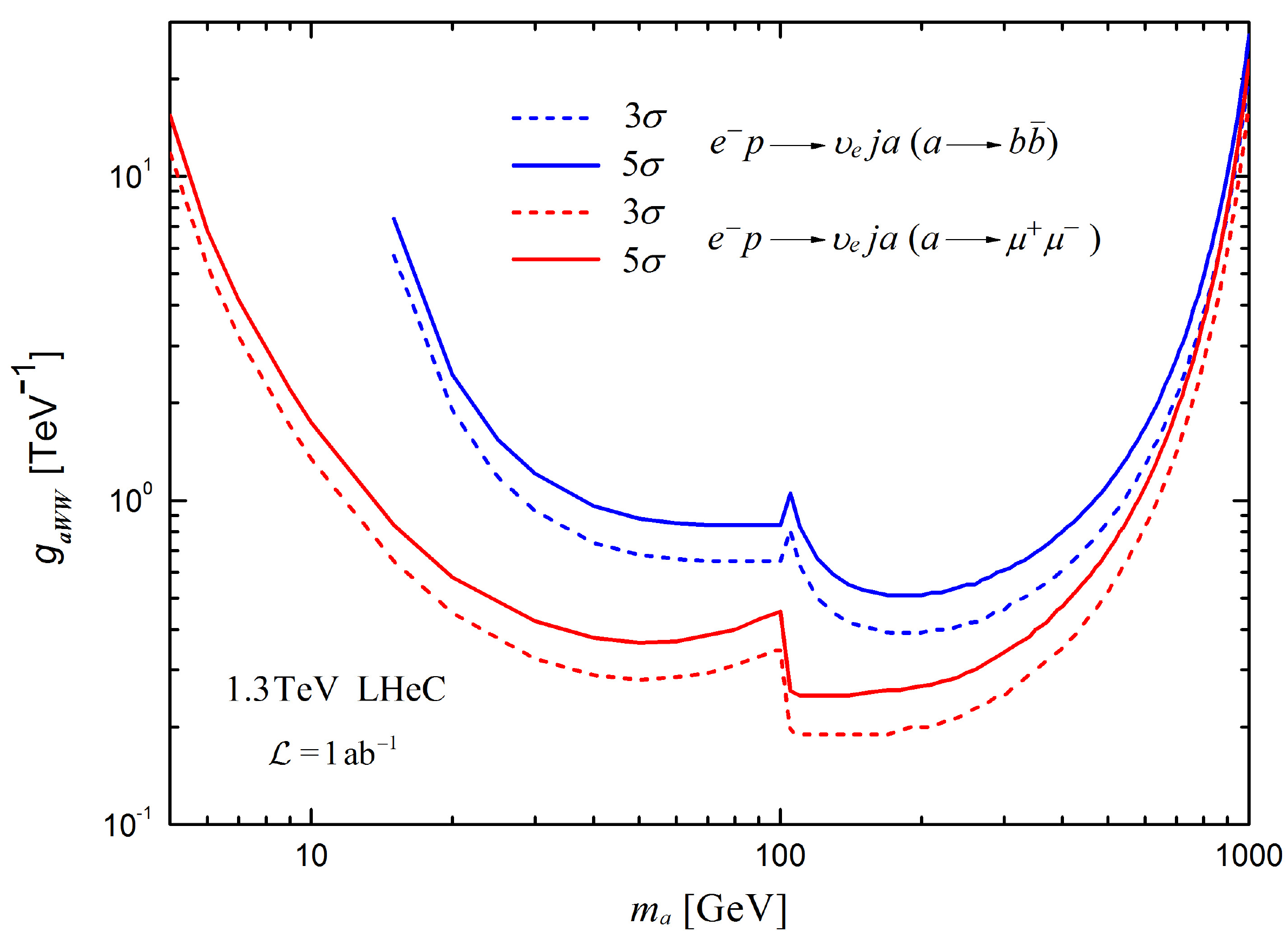}
\caption{The $3\sigma$ and $5\sigma$ curves in the $m _a - g_{aWW}$ plane for the $W^{+}W^{-}$ fusion  processes $e^{-}p\rightarrow\nu_{e}ja(a\rightarrow\mu^{+}\mu^{-})$ (red) and $e^{-}p\rightarrow\nu_{e}ja(a\rightarrow b\overline{b})$ (blue) at the 1.3 TeV LHeC with $\mathcal{L}=$ $1$ ab$^{-1}$.}
\label{35sigma}
\end{center}
\end{figure}

In order to make the analysis more simplified and the cross sections of the signal processes larger, we study the above two decay channels by taking the branching ratios to be $100\%$ for $a\rightarrow\mu^{+}\mu^{-}$  and $a\rightarrow b\overline{b}$, respectively. Furthermore, we plot the $3\sigma$ and $5\sigma$ curves in the plane of $m _a - g_{aWW}$ at the LHeC with $\sqrt{s}=1.3$ TeV and $\mathcal{L}=$ $1$ ab$^{-1}$ in FIG.~\ref{35sigma}, where the expected bounds for the signal processes $e^{-}p\rightarrow\nu_{e}ja(a\rightarrow\mu^{+}\mu^{-})$ and $e^{-}p\rightarrow\nu_{e}ja(a\rightarrow b\overline{b})$ are shown in red and blue, respectively. The figure indicates that the sensitivity bounds on the coupling coefficient $g_{aWW}$ can reach $0.19$ TeV$^{-1}$ ($0.25$ TeV$^{-1}$) and $0.39$ TeV$^{-1}$ ($0.51$ TeV$^{-1}$) at $3\sigma$ ($5\sigma$) confidence level for the dacay channels $a\rightarrow\mu^{+}\mu^{-}$ and $a\rightarrow b\overline{b}$, respectivly. Therefore, the signal of ALP might be probed via the $W^{+}W^{-}$ fusion processes $e^{-}p\rightarrow\nu_{e}ja(a\rightarrow\mu^{+}\mu^{-})$ and $e^{-}p\rightarrow\nu_{e}ja(a\rightarrow b\overline{b})$ at the LHeC, of which the former is more sensitive to search for the ALP in the studied mass range. If the values for the branching ratios of ALP decaying into specific final states are not fixed, the sensitivity obtained by the $bb$ mode will be stronger than that achieved from the $\mu^{+}\mu^{-}$ mode under study due to the couplings of ALP to fermions proportional to the corresponding fermion mass, which results in the signal cross sections of the $bb$ mode being larger than those of the $\mu^{+}\mu^{-}$ mode. We will comprehensively investigate the ALP signal for this case in future work.

\section{Summary and discussion}

Owing to the large energy, the enhanced luminosity and the cleanliness of the hadronic final states, the LHeC will scrutinise the SM deeper than ever before and has its own potential to discover new physics. Exotic phenomena can be studied at the LHeC. With a wide range of phenomenological applications, ALP provides a well-motivated new physics scenario. In this paper, our attention was focused on discussing the possibility of detecting ALP through the $W^{+}W^{-}$ fusion processes $e^{-}p\rightarrow\nu_{e}ja(a\rightarrow\mu^{+}\mu^{-})$ and $e^{-}p\rightarrow\nu_{e}ja(a\rightarrow b\overline{b})$ in the reasonable parameter space at the $1.3$ TeV LHeC with $\mathcal{L}=$ $1$ ab$^{-1}$.

\begin{figure}[H]
\begin{center}
\centering\includegraphics [scale=0.42] {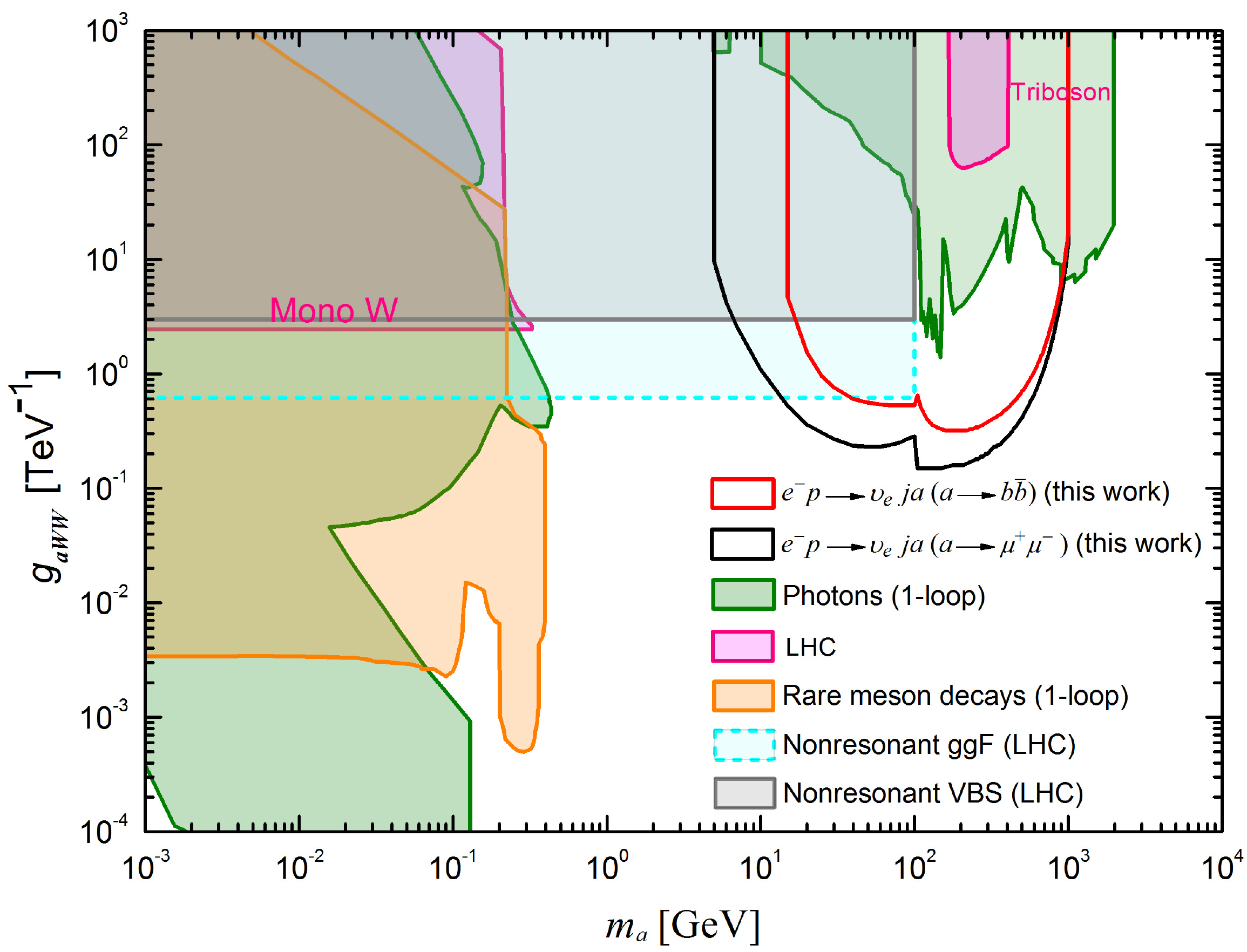}
\caption{Our projected $2\sigma$ sensitivity limits on the coupling of ALP to $W^{\pm}$ bosons at the LHeC in comparison with other current excluded regions.}
\label{ma-gaWW-plane}
\end{center}
\end{figure}

The sensitivity limits on the coupling of ALP to $W^{\pm}$ bosons at $95\%$ C.L. derived in this paper and other exclusion regions from previous studies are given in FIG.~\ref{ma-gaWW-plane}. Our results are shown in black line indicating the decay channel $a\rightarrow\mu^{+}\mu^{-}$ and red line for the decay channel $a\rightarrow b\overline{b}$. The promising sensitivities of the coupling coefficient $g_{aWW}$ are expected to be $0.15\sim6.66$ TeV$^{-1}$ for the signal process $e^{-}p\rightarrow\nu_{e}ja(a\rightarrow\mu^{+}\mu^{-})$ in the ALP mass interval $14\sim924$ GeV at the LHeC with $\sqrt{s}=1.3$ TeV and $\mathcal{L}=$ $1$ ab$^{-1}$. In the case of the signal process $e^{-}p\rightarrow\nu_{e}ja(a\rightarrow b\overline{b})$, the values of $g_{aWW}$ as $0.32\sim6.67$ TeV$^{-1}$ corresponding to the ALP mass interval $39\sim900$ GeV are acquired.

Other regions excluded for the coupling of ALP to $W^{\pm}$ bosons are then depicted in FIG.~\ref{ma-gaWW-plane} as coloured areas. Since the radiative corrections of the ALP-boson couplings to the coupling of ALP with photons, the bounds on ALP-photon coupling can be translated into the bounds for $g_{aWW}$, which are labeled as ``Photons". Such constraints depend mildly on $f_a$, stemming from Beam Dump experiments, supernova SN1987a observations and the LHC results~\cite{Alonso-Alvarez:2018irt}. Limits labeled as ``LHC" are derived from mono-$W$ final states in the sub-GeV ALP mass region~\cite{Brivio:2017ije} and resonant triboson searches as the ALP mass near the TeV range at the LHC~\cite{Craig:2018kne}, though they are superseded by ``Photons" exclusions. Light ALPs can also be produced from rare meson decays and the coupling of ALP to $W^{\pm}$ bosons is best tested by its one-loop impact on rare meson decay experiments, which gives rise to the constraints shown with the orange area in FIG.~\ref{ma-gaWW-plane}~\cite{BNL-E949:2009dza,Izaguirre:2016dfi}. The constraints on $g_{aWW}$ offered by the search for ALP inducing nonresonant signals at the LHC  via gluon-gluon fusion process (labeled ``Nonresonant ggF") and vector boson scattering (labeled ``Nonresonant VBS") processes have been studied in Refs.~\cite{Carra:2021ycg} and~\cite{Bonilla:2022pxu}, respectively. Nevertheless, the ``Nonresonant ggF" bounds can be completely lifted as the interaction of ALP with gluons tends to zero. As a whole, our results cover regions of the parameter space that complement the research of $g_{aWW}$ by the LHC. The obtained LHeC sensitivity limits are stronger than the bounds given by ``Nonresonant ggF" in the cases of $m _a > 14$ GeV for the decay channel $a\rightarrow\mu^{+}\mu^{-}$ and $m _a > 39$ GeV for the $a\rightarrow b\overline{b}$ production mode. The parameter regions being explored are excluded by the ``Photons" constraints for approximately $m _a > 900$ GeV.

In conclusion, comparing with the regions excluded by other experiments, the LHeC is more sensitive to the coupling of ALP with $W^{\pm}$ bosons via the $W^{+}W^{-}$ fusion processes $e^{-}p\rightarrow\nu_{e}ja(a\rightarrow\mu^{+}\mu^{-})$ and $e^{-}p\rightarrow\nu_{e}ja(a\rightarrow b\overline{b})$ for the ALP with mass range of roughly a few tens of GeV to $900$ GeV. The LHeC running at $1.3$ TeV with the integrated luminosity of $1$ ab$^{-1}$ would have great potential in detecting ALPs.

\section*{ACKNOWLEDGMENT}

This work was partially supported by the National Natural Science Foundation of China under Grant No. 11875157 and No. 12147214.


\bibliography{ALPref}

\end{document}